\documentclass{article}
\usepackage{microtype}
\usepackage{graphicx}
\usepackage{subfigure}
\usepackage{booktabs}
\usepackage{hyperref}
\usepackage{multirow}
\usepackage{amssymb}

\usepackage[accepted]{icml2021}

\icmltitlerunning{How Robust are Limit Order Book Representations under Data Perturbation?}

\begin{document}

    \twocolumn[
        \icmltitle{How Robust are Limit Order Book Representations \\ under Data Perturbation?}
        
        \begin{icmlauthorlist}
            \icmlauthor{Yufei Wu}{jpmldn}
            \icmlauthor{Mahmoud Mahfouz}{jpmldn,imp}
            \icmlauthor{Daniele Magazzeni}{jpmldn}
            \icmlauthor{Manuela Veloso}{jpmny}
        \end{icmlauthorlist}
        
        \icmlaffiliation{jpmldn}{J.P. Morgan AI Research, London, United Kingdom}
        \icmlaffiliation{jpmny}{J.P. Morgan AI Research, New York, United States of America}
        \icmlaffiliation{imp}{Imperial College London, London, United Kingdom}
        
        \icmlcorrespondingauthor{Yufei Wu}{yufei.wu@jpmorgan.com}
        \icmlkeywords{Limit Order Books, Deep Learning, Market Microstructure}
        \vskip 0.3in
    ]
    
    \printAffiliationsAndNotice{}
    
    \begin{abstract}
    The success of machine learning models in the financial domain is highly reliant on the quality of the data representation. In this paper, we focus on the representation of limit order book data and discuss the opportunities and challenges for learning representations of such data. We also experimentally analyse the issues associated with existing representations and present a guideline for future research in this area.
\end{abstract}
    \section{Introduction} \label{introduction}

\begin{figure}[!htb]
        \centering
        \includegraphics[width=0.45\textwidth]{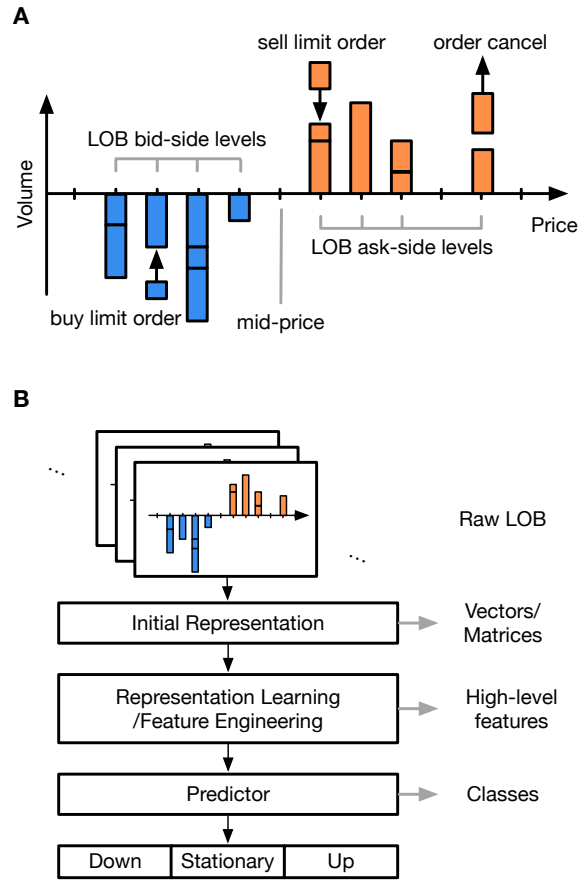}
        \caption{(\textbf{A}) A snapshot of the limit order book. (\textbf{B}) Workflow of a price forecasting task using LOB data with machine learning models.}
        \label{fig:lob}
    \end{figure}

    The limit order book (LOB) is used by financial exchanges to match buyers and sellers of a particular instrument and acts as an indicator of the supply and demand at a given point in time. It can be described as a self-evolving process with complex spatial and temporal structures revealing the price dynamics at the microstructural level. Market making, optimal execution and statistical arbitrage strategies, all require a good understanding of the LOB and its dynamics. Figure \ref{fig:lob} (A) shows a snapshot of the LOB with both the bid (buy) and ask (sell) order volumes accumulated at each price level. The mid-price is the average of the best (lowest) ask price and the best (highest) bid price and the difference between them is referred to as the bid-ask spread. The LOB gets updated continuously with order placements, cancellations and executions.
    
    \begin{figure*}[!tb]
        \centering
        \includegraphics[width=\textwidth]{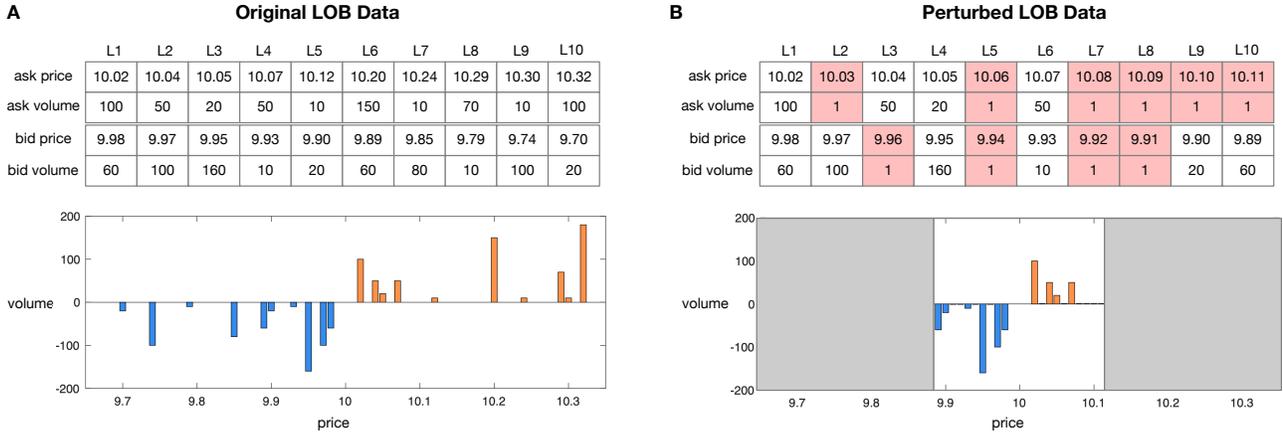}
        \caption{(\textbf{A}) Original LOB data with 10 levels on ask and bid side without perturbation. (\textbf{B}) LOB data with 10 levels after data perturbation. Red blocks represent intentionally placed perturbation orders with order volume = 1. Compared with the original one, the new 10-level data representation has a much narrower vision on the market.}
        \label{fig:perturbation}
    \end{figure*}
    
    The use of algorithmic trading strategies and the digitisation of exchange activities has made available a tremendous amount of LOB data for practitioners and researchers to study the market dynamics from data-driven approaches. This led to a surge in interest for big data applications in the financial markets and machine learning (including deep learning) models becoming a trend in the quantitative finance domain \cite{buehler2019deep}, \cite{wiese2020quant}. The LOB data come in different degrees of granularity with \emph{L1} data providing the best bid/ask prices and volumes, \emph{L2} data providing the same data across all price levels and \emph{L3} data containing the non-aggregated orders placed by market participants.
    
    In our work, we focus on how LOB data is typically represented by taking a price forecasting task as an example. In practice, a vector representation of the raw limit order book information is needed for upcoming learning processes. This transformation from raw data to feature vectors is typically referred to as \emph{feature engineering}, which requires a good and comprehensive understanding of the domain knowledge to make sure the extracted features match the learning task. By contrast, \emph{representation learning}, also called \emph{feature learning}, is an automated approach to discover an optimal representation for the data. The major difference between feature engineering and representation learning is whether the representation is formed in a purely data-driven way. Also, it is common for a machine learning system to involve both feature engineering and representation learning with multiple levels of representation appearing at different stage of processing (see figure \ref{fig:lob} (B)). 
    
    The performance of machine learning models is heavily influenced by the data representation scheme \cite{bengio2013representation}. For neural networks, the representation learning and the prediction processes are combined within the network structure and are trained together towards the same target function. In this case, the original representation of LOB, \textit{i.e.} the input representation to neural networks, becomes the foundation of the entire model. Presently, the price level-based data representation scheme is used in almost all recent studies \cite{tsantekidis2017using,tsantekidis2017forecasting,tran2018temporal,zhang2019deeplob,mahfouz2019importance,sirignano2019deep,tsantekidis2020using,wallbridge2020transformers} applying deep learning models on LOB data. However, this representation scheme is rarely discussed or investigated towards its compatibility with machine learning especially deep learning models. In this paper, we propose a pioneer insight to challenge this level-based LOB representation for machine learning models, by showing potential risks under subtle perturbations and raising concerns regarding to its robustness. 
    
    \paragraph{Summary of contribution} This paper propose a perturbation paradigm to the LOB. By examining the performance change of LOB price forecasting machine learning models under perturbation, we examine the robustness of data representation. The experimental results confirm our concerns about the current level-based LOB representation as well as machine learning models designed based on this representation scheme. Furthermore, this paper present desiderata of LOB representations for guiding future research in this area.

    \section{Problem Description} \label{problem_description}
    
    We first introduce the commonly-used level-based LOB representation scheme found in benchmark and research datasets (e.g. \cite{ntakaris2018benchmark,huang2011lobster}). The spatial representation of this scheme, \textit{i.e.} an LOB snapshot, is a vector $s_{t} = \left\{p_{a}^{i}(t), v_{a}^{i}(t), p_{b}^{i}(t), v_{b}^{i}(t)\right\}_{i=1}^{L}$, where, $p_{a}^{i}(t)$, $p_{b}^{i}(t)$ are the ask and bid prices for price level $i$ and $v_{a}^{i}(t)$, $v_{b}^{i}(t)$ are the ask and bid volumes respectively. Temporally, the history of the LOB snapshots is stacked to reflect an evolution of the market, leading to the commonly-used limit order book data structure $S\in \mathbb{R}^{T\times 4L}$ where T is the history length and L is the amount of price levels considered for each side. 
    
    This level-based representation is efficient and convenient from the perspective of human understanding and how the matching engine in exchanges works. From the machine learning perspective, this representation has some particular characteristics. The most intuitive one is that the price and volume for each LOB level are tied together - any disentanglement or distortion to this would result in an invalid representation. In addition, the spatial structure across different levels is not homogeneous since there is no assumption for adjacent price levels to have fixed intervals. Note that, homogeneous spatial relationship is a basic assumption for convolutional neural networks (CNN) due to the parameter sharing mechanism. Thus, the heterogeneous spatial feature of level-based LOB data may reduce model robustness when learning with CNN models. From the temporal perspective, we also realise some instability of the representation due to occasional shifts of price levels - the previous best bid/ask data can suddenly shift to second best bid/ask channel if a new order is placed with a better price.
    

    \section{Data Perturbation} \label{methods}
    
    We present a simple data perturbation method to examine the robustness of the price level-based representation from the machine learning perspective. In some LOB data for equities, the price difference between adjacent price levels is sometimes larger than the tick size (the minimum price increment change allowed). This is especially prevalent in small-tick stocks and can result in the entire LOB shifting even if a small order of the minimum allowable size is placed at a price in between the existing price levels. The data perturbation method presented assumes that the data is perturbed by small size orders at empty price levels beyond the best ask/bid prices. This perturbation ensures no change is made to the mid-price before and after perturbation to make sure the prediction labels are not affected. 
    
    We illustrate this data perturbation with a synthetic LOB example as shown in Fig.  \ref{fig:perturbation}. Fig. \ref{fig:perturbation} (A) shows the synthetic LOB snapshot with 10 price levels in both ask and bid sides of the LOB (marked as L1-L10) before any perturbation. We assume the tick size is 0.01 and the minimum order size present in our data is 1. In this LOB snapshot, the mid-price is 10.00 with bid-ask spread equal to 0.04. We can observe some price levels where no orders are placed, such as 10.03, 10.06 in the ask side and 9.96, 9.94 in the bid side. To perturb this LOB data, one can place orders with allowed minimum order size to fill these empty price levels. These minimum size orders may seem to be not influential since 1) they do not effect the mid price, 2) their volumes are tiny. c (see Fig. \ref{fig:perturbation} (B)). Approximately half of the original price level information is no longer visible after perturbation (e.g. ask-side L5 to L10 information is not included in representation after perturbation) and while the rest are preserved, they are shifted to different levels in the LOB representation (e.g., the ask-side L2 appears in ask-side L3 after perturbation). 
    
    Intuitively, this perturbation has two impacts from the machine learning point of view. Firstly, it shifts the 40-dimensional input space dramatically. For example, the Euclidean distance between these two 40-dimensional vectors before and after perturbation is 344.623 whereas actually the total volume of orders applied is only 10. This means that the level-based representation scheme does not bring local smoothness. Furthermore, it narrows the scope of vision of machine learning models to `observe' the market. As shown in the LOB data visualisation plot in Fig. \ref{fig:perturbation}, the gray areas are masked out for the model input after perturbation.
    
    \section{Related Work} \label{related_work}
    
    The study of the importance of robust data representation, the criteria for evaluating the quality of the representation, and the variety of methods for learning these representations is studied extensively in the machine learning literature with \cite{bengio2013representation} providing a survey of these methods. In our work, we focus on the representation of financial market microstructure data. \cite{bouchaud2018trades}, \cite{abergel2016limit} study the structure and empirical properties of limit order books and provide a set of statistical properties (referred to as \textit{stylized facts}) using NASDAQ exchange data. On the other hand, \cite{lehalle2018market} discusses the practical aspects and issues of market structure, design, price formation, discovery and the behaviour of different actors in limit order book markets. A significant amount of research in recent years focused on applying deep learning models on limit order book data for the purposes of price forecasting or price movements classification. Different model architectures were investigated including multi-layer perceptrons \cite{mahfouz2019importance}, recurrent neural networks \cite{sirignano2019universal}, convolutional neural networks \cite{zhang2019deeplob}, \cite{zhang2018bdlob} and self-attention transformer networks \cite{wallbridge2020transformers}.
    \section{Experiments and Results} \label{results}
    
    In this section, we implement a series of experiments to examine the robustness of the LOB representation under different data perturbations. We take price forecasting as the task paradigm and train various machine learning models to perform such task and examine their performance when encountering unexpected perturbation.

    \subsection{Benchmark Dataset and Models}
    
        We use the FI-2010 dataset \cite{ntakaris2018benchmark} as the benchmark dataset. The FI-2010 dataset consists of LOB data from 5 stocks in the Helsinki Stock Exchange during normal trading hours (no auctions) for 10 trading days. This dataset takes into account 10 price levels on the bid/ask sides of the limit order book, which are updated according to events such as order placement, executions and cancellations. In our experiments, we take into account the history of the LOB snapshots for future price movement prediction. Thus, each input data point is a short time series with input dimension $T \times 40$ where T is the total amount of historical snapshots. For experiments in this paper, we choose T = 10. 
        
        The prediction target is the micro-movement $l_t = \frac{m_+(t)-p_t}{p_t}$ where $m_+(t) = \frac{1}{k} \sum^k_{i=1}p_{t+i}$ is the smoothed mid-price with prediction horizon $k$. The movement is further categorised into three classes - 1:up ($l_t>0.002$), 2:stationary ($-0.002<l_t<0.002$), 3:up ($l_t<-0.002$). We choose the FI-2010 prediction labels with prediction horizon k-50 as the targets for model training and testing. The training set is relatively balanced with 34\%/ 32\%/ 34\% components of labels Up / Stationary / Down samples. By contrast, the testing set is an unbalanced set with 28\%/ 47\%/ 25\% samples for Up / Stationary / Down. 
        
        We choose 4 benchmark price forecasting models, \textit{i.e.} logistic regression, multi-layer perceptron (MLP), long short term memory (LSTM) and the DeepLOB model combining convolutional networks with LSTM \cite{zhang2019deeplob}. All these methods are trained with the same FI-2010 training dataset and then tested in 4 types of scenarios: the data is not perturbed (`None') and the data is perturbed by placing minimum-size orders to fill the ticks on the ask side only (`ask-side'), on the bid side only (`bid-side'), on both the ask and bid side. 
        
    \subsection{Model Performance under Perturbation}
    
        Table \ref{table:performance} demonstrates the testing performance of these machine learning models in the price movement forecasting tasks. Since the testing set is unbalanced, we use 4 different metrics (scores) to evaluate and compare the performance - Accuracy (\%), Precision (\%), Recall (\%) and F-score (\%). Among these metrics, Accuracy (\%) is measured as the percentage of predictions of the test samples exactly matches the ground truth, which is the unbalanced accuracy score where as the rest metrics are all averaged across classes in an unweighted manner to eliminate the influence of data imbalance. 
        
        We observe a performance decay of all the machine learning models under the unexpected data perturbation introduced, especially these sophisticated models like LSTM and DeepLOB. In addition, the performance decline alters under different types of data perturbation. Compared with no perturbation scenario, ask-side and bid-side perturbations cause around 6\% accuracy decrease on MLP, 7\% on LSTM and 9\%-14\% on DeepLOB. When the perturbation is applied to both sides, the performance decrease becomes more severe - 11\% accuracy decrease on MLP, 12\% on LSTM and 30\% on DeepLOB. Similar trends can also be viewed for other evaluation metrics. 
        
        From the these performance decay results, we find that DeepLOB, the best performed model under normal condition as well as the most complicated one, is also the most vulnerable one under perturbation. Its predictive accuracy decreases to 47.5\% and the F-score is only 22.2\%, which even underperforms logistic regression. The reason behind this phenomenon may be a combination of various factors. On one hand, the complexity of model is related to overfitting, which may reduce the generalisation ability and become unstable under the perturbation. Also as we mentioned in earlier sections, CNN assumes homogeneous spatial relationship but the level-based LOB representation is obviously heterogeneous, which leads to a mismatch between the data representation and the network characteristics. Once the spatial relationship is further broken due to perturbation, the CNN descriptors may not be able to extract meaningful features and thus cause malfunction of the entire predictor. 
    
        \begin{table}[!tb]
            \centering
            \small
            \begin{tabular}{c c c c c } 
                 \toprule 
                 \multirow{2}{*}{Data Perturbation} & 
                 \multicolumn{4}{c}{Metrics (\%)} \\
                 \cmidrule(lr){2-5}
                 & Accuracy  & Precision & Recall  & F-score \\
                 \midrule
                 \multicolumn{5}{c}{Logistic Regression} \\
                 \midrule
                 None & 52.9 & 62.9 & 41.6 & 38.2\\
                 Ask & 49.7 & 50.4 & 42.9 & 42.4\\
                 Bid & 51.5 & 62.2 & 40.3 & 35.7\\
                 Both sides & 49.3 & 45.8 & 42.0 & 41.5\\
                 \midrule
                 \midrule
                 \multicolumn{5}{c}{MLP} \\
                 \midrule
                 None & 61.1 & 66.6 & 54.7 & 55.6\\
                 Ask & 55.5 & 61.4 & 49.2 & 48.5\\
                 Bid & 56.5 & 63.6 & 46.8 & 46.1\\
                 Both sides & 51.0 & 50.0 & 41.5 & 39.3\\
                 \midrule
                 \midrule
                 \multicolumn{5}{c}{LSTM} \\
                 \midrule
                 None & 70.4 & 68.4 & 68.0 & 68.2 \\
                 Ask & 63.6 & 62.1 & 62.0 & 61.5\\
                 Bid & 63.0 & 61.2 & 61.0 & 60.5\\
                 Both sides & 57.9 & 54.7 & 54.9 & 54.8\\
                 \midrule
                 \midrule
                 \multicolumn{5}{c}{DeepLOB \cite{zhang2019deeplob}} \\
                 \midrule
                 None & 77.2 & 76.2 & 74.3 & 75.1 \\
                 Ask & 68.0 & 70.1 & 60.8 & 62.3 \\
                 Bid & 63.2 & 68.6 & 54.6 & 55.5 \\
                 Both sides & 47.5 & 47.9 & 33.7 & 22.2 \\
                 \bottomrule
            \end{tabular}
            \caption{Price forecasting model performance under data perturbation. Each model is trained with a non-perturbed training set and when testing the model, we apply various data perturbation. None: no perturbation. Ask-side: perturbation only applied to the ask-side of data. Bid-side: perturbation only applied to the bid-side of the data. Both sides: perturbation applied to both ask and bid sides.} 
            \label{table:performance}
        \end{table}
        \begin{figure*}[!tb]
        \centering
        \includegraphics[width=0.95\textwidth]{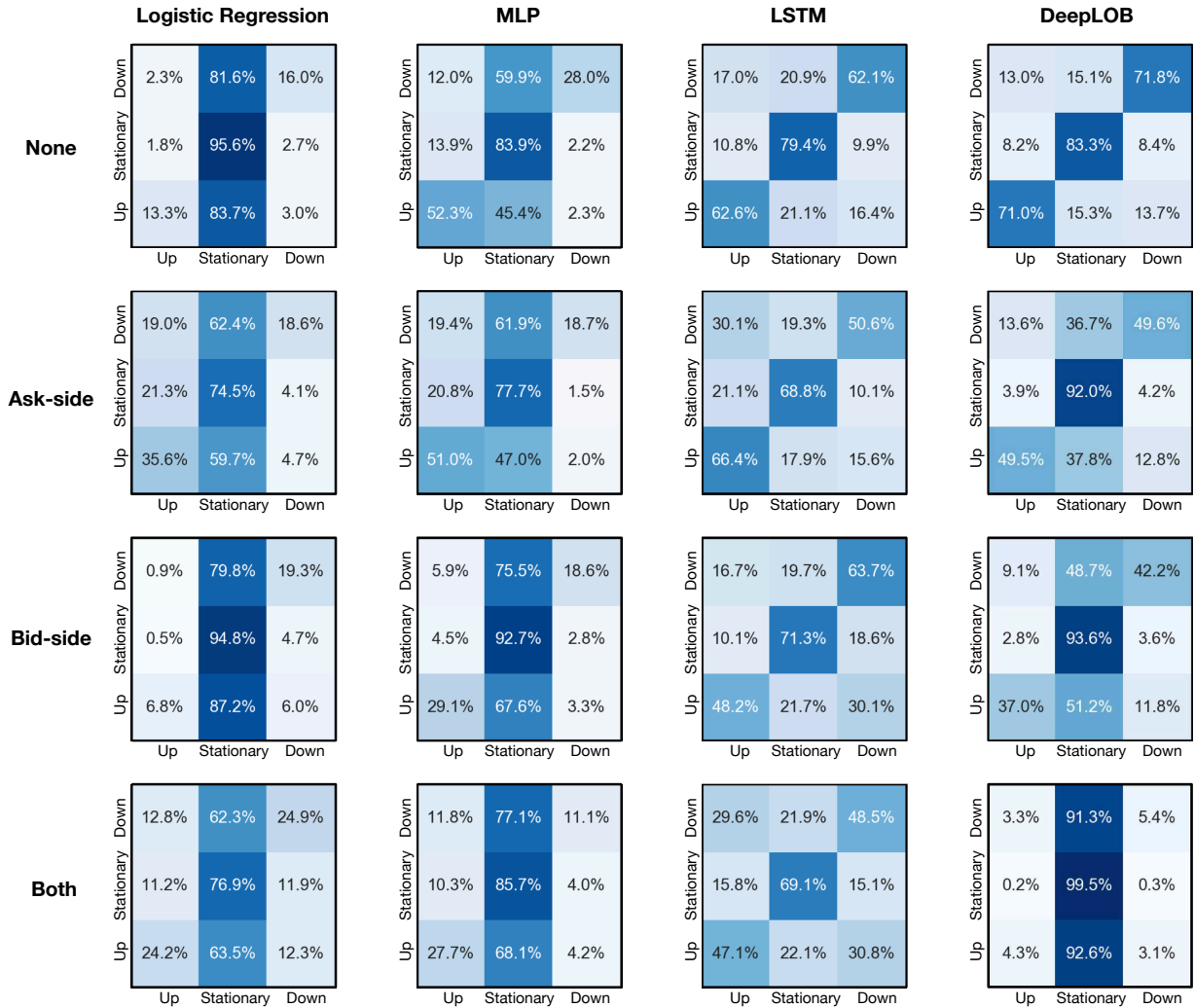}
        \caption{Confusion matrices for corresponding experimental results in Table. \ref{table:performance}}
        \label{fig:cm}
    \end{figure*}
    
        Fig. \ref{fig:cm} further illustrates more details behind the numerical performance metrics in the form of a confusion matrix. The logistic regression model basically classify a majority of samples as `Stationary' no matter whether perturbation is applied. Similarly in MLP, about half of the `Up' and `Down' samples are misclassified as `Stationary' ones. Both LSTM and DeepLOB shows confusion matrices with obvious diagonal feature without perturbation - more than half of the samples from each class are classified the same as their true labels. When the perturbation is applied, LSTM shows performance decrease but the still near half of samples are correctly classified. DeepLOB, however, fails in the perturbation condition by misclassifying almost all the data to `Stationary' class (see DeepLOB+Both in Fig. \ref{fig:cm}).
            
    \section{Conclusion and Future Work} \label{conclusion}

In this paper, we discussed the importance of data representations to machine learning models applied to LOB-related tasks. We designed data perturbation scenarios to test the robustness of commonly-used machine learning models with the level-based LOB representation scheme in price forecasting tasks. We show that, although the perturbation is subtle, it still shows a large impact to the level-based representation and thus leads to significant performance decay. In particular, this performance decay is more severe in sophisticated machine learning models. 

Based on the findings in this paper and our understanding of representations, we would like to raise some desiderata for improving the robustness of LOB-related data representations and machine learning models designed on top of certain representations:

\begin{itemize}
    \item \textbf{Smoothness}: LOB Data representations should not change dramatically under subtle perturbations. 
    \item \textbf{Efficiency}: LOB data representation should organise data in a efficient manner to reduce the \emph{curse of dimensionality}. 
    \item \textbf{Stochasticity}: LOB data representation should consider the stochasticity of the market, instead of treating the market process as a deterministic one. 
    \item \textbf{Validity}: Basic assumptions needs to be matched between data representations and learning models. If not, these models may contain unknown risks due to invalid fundamental settings.
\end{itemize} 

Our future work would focus on feature engineering and representation learning for LOB data that can fulfill these desiderata and combining LOB representations with various machine learning tasks including forecasting, reinforcement learning and etc.
    
    \section*{Acknowledgements}
        The authors would like to acknowledge our colleagues Vacslav Gluckov, Rui Silva, Thomas Spooner and Jeremy Turiel and  for their input and suggestions at various key stages of the research.
        
        This paper was prepared for informational purposes by the Artificial Intelligence Research group of JPMorgan Chase \& Co and its affiliates ("J.P. Morgan"), and is not a product of the Research Department of J.P. Morgan. J.P. Morgan makes no representation and warranty whatsoever and disclaims all liability, for the completeness, accuracy or reliability of the information contained herein. This document is not intended as investment research or investment advice, or a recommendation, offer or solicitation for the purchase or sale of any security, financial instrument, financial product or service, or to be used in any way for evaluating the merits of participating in any transaction, and shall not constitute a solicitation under any jurisdiction or to any person, if such solicitation under such jurisdiction or to such person would be unlawful.
    
    \bibliography{references}
    \bibliographystyle{icml2021}

\end{document}